\documentclass[12pt,twoside]{article}

\date{} 

\begin{document} 

\centerline{\bf Adv. Studies Theor. Phys., Vol. 4, 2010, no. 20, 945 - 949.} 

\centerline{} 

\centerline{} 

\centerline {\Large{\bf CHSH and local hidden causality}} 

\centerline{} 


\centerline{} 

\centerline{\bf {J.F. Geurdes}} 

\centerline{} 

\centerline{C. vd. Lijnstraat 164 2593 NN Den Haag Netherlands} 

\centerline{han.geurdes@gmail.com} 



\centerline{} 

\begin{abstract} Mathematics equivalent to Bell's derivation of the inequalities, also allows a local hidden variables explanation for the correlation between distant measurements.
\end{abstract} 

{\bf Keywords:} Bell inequalities, classical probability, EPR paradox.

\section{Introduction} 
Bell inequalities \cite{Bell} are a well studied subject. To many the experimental verification of the violation of inequalities e.g. \cite{Asp}, \cite{Weihs} is sufficient evidence for the completeness of quantum theory.  Here, it will be demonstrated that Bell's form of local hidden correlation
\begin{equation}\label{e1}
P(\vec{a},\vec{b}) = \int\limits_{\lambda \in \Lambda} \rho_{\lambda} A_{\lambda}(\vec{a})B_{\lambda}(\vec{b}) d \lambda
\end{equation}
can be transformed to violate Bell's inequality. We have, $\vec{a}$ and $\vec{b}$ for unitary parameter vectors of e.g. Stern-Gerlach magnets in an ortho-positronium decay experiment. $\lambda$ represents the extra hidden parameters in  a set $\Lambda$. The probability density $\rho_{\lambda}$ is a classical density. The measurement functions $ A_{\lambda}(\vec{a})$ and $B_{\lambda}(\vec{b})$ project in $\{-1 ,1\}$.  Bell showed, using the expression below, that models with a classical probability density may not violate the inequality\footnote{ If there is no confusion the $d\lambda$ will be suppressed.}.
\begin{equation}\label{e2}
P(\vec{a},\vec{b}) - P(\vec{x},\vec{y}) =  \int\limits_{\lambda \in \Lambda} \rho_{\lambda} A_{\lambda}(\vec{a})B_{\lambda}(\vec{b})A_{\lambda}(\vec{x})B_{\lambda}(\vec{y})\left\{A_{\lambda}(\vec{x})B_{\lambda}(\vec{y})-A_{\lambda}(\vec{a})B_{\lambda}(\vec{b})\right\}
\end{equation}
\subsection{Singlet state Bell inequality}
Bell expressed the singlet state of the electron and positron in the positronium as $\forall: \vec{a} (\left| \vec{a} \right|=1) \forall: \lambda (\lambda \in \Lambda)\left\{A_{\lambda}(\vec{a})+B_{\lambda}(\vec{a})=0\right\}$. The following steps are elementary. Let us take, $\vec{x}=\vec{b}$ and $\vec{y}=\vec{c}$.  With the singlet, we see that equation (\ref{e2}) can be written as
\begin{equation}\label{e3}
P(\vec{a},\vec{b}) - P(\vec{b},\vec{c}) =  \int\limits_{\lambda \in \Lambda} \rho_{\lambda} \left\{A_{\lambda}(\vec{b})A_{\lambda}(\vec{c})-A_{\lambda}(\vec{a})A_{\lambda}(\vec{b})\right\}
\end{equation}
Or, noting $1-A_{\lambda}(\vec{a})A_{\lambda}(\vec{c}) \geq 0$,
\begin{equation}\label{e4}
\left|P(\vec{a},\vec{b}) - P(\vec{b},\vec{c})\right| \leq  \int\limits_{\lambda \in \Lambda} \rho_{\lambda} \left|A_{\lambda}(\vec{c})A_{\lambda}(\vec{b})\right|\left\{1-A_{\lambda}(\vec{a})A_{\lambda}(\vec{c})\right\}
\end{equation}
Because, $\left|A_{\lambda}(\vec{c})A_{\lambda}(\vec{b})\right|=1$ and $\rho_{\lambda}$ classical, we have the Bell inequality
\begin{equation}\label{e5}
\left|P(\vec{a},\vec{b}) - P(\vec{b},\vec{c})\right| \leq 1+ P(\vec{a},\vec{c})
\end{equation}
The quantum correlation is: $P_{qm}(\vec{x}, \vec{y})=-\left(\vec{x}\cdot\vec{y}\right)$. If in two-dimensions, 
$\vec{a}=\left( {-1\over{\sqrt{2}}}, {1\over{\sqrt{2}}} \right) $, $\vec{b}=\left( {1\over{\sqrt{2}}}, {1\over{\sqrt{2}}} \right) $ and $\vec{c}=\left( 0, 1 \right)$, then,  inequality is violated because, $\left| 0 - {-1\over{\sqrt{2}}} \right| \leq 1 - {1\over{\sqrt{2}}}$ is false. Associated to this inequality in equation(\ref{e5}) a more general inequality, the CHSH inequality \cite{Clau}, exists. The principle is the same.
\section{Sets and Integrals}
Keeping an eye on equation (\ref{e2}), hidden parameters sets can be defined
\begin{equation}\label{e6}
\Omega_{\pm}=\left\{ \lambda \in \Lambda | A_{\lambda}(\vec{a})B_{\lambda}(\vec{b})=A_{\lambda}(\vec{x})B_{\lambda}(\vec{y})=\pm 1 \right\}
\end{equation}
and
\begin{equation}\label{e7}
\Omega_{0}=\left\{ \lambda \in \Lambda | A_{\lambda}(\vec{a})B_{\lambda}(\vec{b})=-A_{\lambda}(\vec{x})B_{\lambda}(\vec{y})=\pm 1 \right\}
\end{equation}
Given, $\vec{a}$, $\vec{b}$, $\vec{x}$ and $\vec{y}$, either, $A_{\lambda}(\vec{a})B_{\lambda}(\vec{b})=A_{\lambda}(\vec{x})B_{\lambda}(\vec{y})$ or $A_{\lambda}(\vec{a})B_{\lambda}(\vec{b})=-A_{\lambda}(\vec{x})B_{\lambda}(\vec{y})$ for arbitrary, $\lambda \in \Lambda$. Moreover, $A_{\lambda}(\vec{a})B_{\lambda}(\vec{b})=\pm 1$ for arbitrary, $\lambda \in \Lambda$. Hence, $\Lambda = \Omega_0 \cup \Omega_{+} \cup \Omega_{-}$ and equation (\ref{e2}) is
\begin{equation}\label{e8}
P(\vec{a},\vec{b}) - P(\vec{x},\vec{y}) =  \int\limits_{\lambda \in \Omega_0} \rho_{\lambda} A_{\lambda}(\vec{a})B_{\lambda}(\vec{b})A_{\lambda}(\vec{x})B_{\lambda}(\vec{y})\left\{A_{\lambda}(\vec{x})B_{\lambda}(\vec{y})-A_{\lambda}(\vec{a})B_{\lambda}(\vec{b})\right\}
\end{equation}
From $\Omega_0$ folows $A_{\lambda}(\vec{a})B_{\lambda}(\vec{b})A_{\lambda}(\vec{x})B_{\lambda}(\vec{y})=-1$ and $\left\{A_{\lambda}(\vec{x})B_{\lambda}(\vec{y})-A_{\lambda}(\vec{a})B_{\lambda}(\vec{b})\right\}=2A_{\lambda}(\vec{x})B_{\lambda}(\vec{y})$. Hence,
\begin{equation}\label{e9}
P(\vec{a},\vec{b}) - P(\vec{x},\vec{y}) = -2 \int\limits_{\lambda \in \Omega_0}\rho_{\lambda}A_{\lambda}(\vec{x})B_{\lambda}(\vec{y})
\end{equation}
Suppose, $P(\vec{a},\vec{b})=0$, as 'starting position' in the experiment. This gives a reformulation of $P(\vec{x},\vec{y})$ where $\vec{x}$ and $\vec{y}$ are different form $\vec{a}$ and $\vec{b}$. Hence,
\begin{equation}\label{e10}
P(\vec{x},\vec{y}) = 2 \int\limits_{{\lambda \in \Omega_{0|P(\vec{a},\vec{b})=0}}} \rho_{\lambda}A_{\lambda}(\vec{x})B_{\lambda}(\vec{y})
\end{equation}
Note that according to equation (\ref{e1}) and the $\Omega$ sets we may write for $P(\vec{a},\vec{b})=0$
\begin{equation}\label{e11}
P(\vec{a},\vec{b})=0=\int\limits_{{\lambda \in \Omega_{0|P(\vec{a},\vec{b})=0}}} \rho_{\lambda}A_{\lambda}(\vec{a})B_{\lambda}(\vec{b})+\int\limits_{{\lambda \in \Omega_{+|P(\vec{a},\vec{b})=0}}} \rho_{\lambda}-\int\limits_{{\lambda \in \Omega_{-|P(\vec{a},\vec{b})=0}}} \rho_{\lambda}
\end{equation} 
Moreover, generally $P(\vec{x},\vec{y}) \neq P(\vec{a},\vec{b})$ which follows from comparing equation (\ref{e10}) with (\ref{e11}). Because, in $\Omega_0$, we see for arbitary $\lambda \in \Omega_0$ that $A_{\lambda}(\vec{a})B_{\lambda}(\vec{b})=-A_{\lambda}(\vec{x})B_{\lambda}(\vec{y})=\pm 1$, it follows from equation (\ref{e11}) that we may rewrite $P(\vec{x},\vec{y})$ as
\begin{equation}\label{e13} 
{1\over{2}}P(\vec{x},\vec{y}) = \int\limits_{{\lambda \in \Omega_{+|P(\vec{a},\vec{b})=0}}} \rho_{\lambda}- \int\limits_{{\lambda \in \Omega_{-|P(\vec{a},\vec{b})=0}}} \rho_{\lambda}
\end{equation}
Equations (\ref{e6}) and (\ref{e7}) show that  the $\Omega$ sets depend on $\vec{a}$, $\vec{b}$, $\vec{x}$ and $\vec{y}$. Given $P(\vec{a},\vec{b})=0$, this fixes the $\vec{a}$ and $\vec{b}$. Hence, $\Omega_{\pm|P(\vec{a},\vec{b})=0}=\Omega_{\pm|P(\vec{a},\vec{b})=0}(\vec{x},\vec{y})$, implicit in equation(\ref{e13}).  Start the experiment with two parameters $\vec{a}$ and $\vec{b}$ that produces the condition $P(\vec{a},\vec{b})=0$ and let $\vec{x}$ and $\vec{y}$ free\footnote {see the discussion section}. $\vec{x}$ does not afect  $B_{\lambda}(\vec{y})$ and vice versa, hence, no locality violation.
\section{Violation CHSH}
We will show that there is a classical probability density that allows violation of the CHSH $\left| D \right|\leq 2$, with,
\begin{equation}\label{e14}
D=P(1_A,1_B)-P(1_A,2_B)-P(2_A,1_B)-P(2_A,2_B)
\end{equation}
Here, $1_{A(B)}$ and $2_{A(B)}$ are unitary vectors randomly selected by $A(B)$.
\subsection{Probability density}
We postulate a density for $(\lambda_1,\lambda_2) \in [{-1\over{\sqrt{2}}},{1\over{\sqrt{2}}}]\times[{-1\over{\sqrt{2}}},{1\over{\sqrt{2}}}] = \Lambda$ with $n=1,2$
\begin{equation}\label{e15}
\rho_{\lambda_n}=\left\{ \begin{array}{lcl}
{{1\over{\sqrt{2}}},~~{-1\over{\sqrt{2}}} \leq \lambda_n \leq {1\over{\sqrt{2}}}} \\
{0,~~~elsewhere}
\end{array}\right.
\end{equation} 
This density is Kolmogorovian.  
\subsection{Selection of parameters}
We establish the parameter vectors that the observers $A$ and $B$ will use. For $A$, $1_A=(1,0)$ and $2_A=(0,1)$. For $B$, $1_B=({1\over{\sqrt{2}}},{-1\over{\sqrt{2}}})$ and $2_B=({-1\over{\sqrt{2}}},{-1\over{\sqrt{2}}})$. If we take the quantum correlation, it follows,  $P_{qm}(1_A,1_B)={-1\over{\sqrt{2}}}$, $P_{qm}(1_A,2_B)={1\over{\sqrt{2}}}$, $P_{qm}(2_A,1_B)={1\over{\sqrt{2}}}$ and $P_{qm}(2_A,2_B)={1\over{\sqrt{2}}}$. Quantum mechanics violates $\left| D \right|\leq 2$,  because $|D|=2\sqrt{2}$ is found.
Because, $\rho_{\lambda_1}\rho_{\lambda_2}={1\over{2}}$ for $(\lambda_1,\lambda_2) \in [{-1\over{\sqrt{2}}},{1\over{\sqrt{2}}}]\times[{-1\over{\sqrt{2}}},{1\over{\sqrt{2}}}]$ and $\Omega_{\pm|P(\vec{a},\vec{b})=0}(\vec{x},\vec{y}) \subset [{-1\over{\sqrt{2}}},{1\over{\sqrt{2}}}]\times[{-1\over{\sqrt{2}}},{1\over{\sqrt{2}}}]$, we obtain from equation (\ref{e13})
\begin{equation}\label{e16}
P(\vec{x},\vec{y}) = \int\limits_{{\lambda \in \Omega_{+|P(\vec{a},\vec{b})=0}}(\vec{x},\vec{y}) } d \lambda_1d \lambda_2 - \int\limits_{{\lambda \in \Omega_{-|P(\vec{a},\vec{b})=0}}(\vec{x},\vec{y}) }d \lambda_1 d \lambda_2
\end{equation}
If, subsequently, observer $A$ selects $1_A$, then the hidden parameter $\lambda_1$ is in $[{-1\over{\sqrt{2}}},1-{1\over{\sqrt{2}}}]\subset [{-1\over{\sqrt{2}}},{1\over{\sqrt{2}}}]$. If, $A$ selects $2_A$ then $\lambda_1$ is in $[-1+{1\over{\sqrt{2}}},{1\over{\sqrt{2}}}]\subset [{-1\over{\sqrt{2}}},{1\over{\sqrt{2}}}]$. Similarly, if $B$ selects $1_B$, then then $\lambda_2$ is in $[0,{1\over{\sqrt{2}}}]\subset [{-1\over{\sqrt{2}}},{1\over{\sqrt{2}}}]$. Finally, if $B$ selects $2_B$, then $\lambda_2$ is found in $[{-1\over{\sqrt{2}}},0]\subset [{-1\over{\sqrt{2}}},{1\over{\sqrt{2}}}]$. The intervals responding to settings do not violate locality: $A$ settings are associated to $\lambda_1$ intervals, $B$ settings to $\lambda_2$ intervals. Suppose $A$ selects $1_A$ and $B$ selects $1_B$. We turn to $\Omega_{\pm|P(\vec{a},\vec{b})=0}({1_A},{1_B})$. If, $\Omega_{+|P(\vec{a},\vec{b})=0}({1_A},{1_B})=\emptyset$ and 
$\Omega_{-|P(\vec{a},\vec{b})=0}({1_A},{1_B})=[{-1\over{\sqrt{2}}},1-{1\over{\sqrt{2}}}] \times [0,{1\over{\sqrt{2}}}]$, from equation (\ref{e16}) it follows that $P(1_A,1_B)={-1\over{\sqrt{2}}}$. Hence, a selection of $\Omega_{\pm|P(\vec{a},\vec{b})=0}(\vec{x},\vec{y})$ is possible giving $|D|>2$. 
\section{Conclusion and discussion}
The result of violating $|D| \leq 2$ with proper $\Omega_{\pm|P(\vec{a},\vec{b})=0}(\vec{x},\vec{y})$ and locality obeying interval selection rules, is surprising.  The mathematics was similar to the one used by Bell \cite{Bell}. Moreover, no violations of locality were introduced. In a random selection experiment there is a non-zero probability that, combined with the deterministic interval selection, a proper selection of $\Omega_{\pm|P(\vec{a},\vec{b})=0}(\vec{x},\vec{y})$ is obtained. When Bell's reasoning is sound, no violation should be possible {\it at all} with the use of classical local hidden models given the employed parameters.  Note that other violating instances can be treated similarly. If there can be no reasons given why locality and causality selections of $\Omega_{\pm|P(\vec{a},\vec{b})=0}(\vec{x},\vec{y})$ are impossible, then a local hidden variable explanation of experiments cannot be excluded. The transformation of (\ref{e1}) is based on a single fixing of $\vec{a}$ and $\vec{b}$, {\it independent} of the $\vec{x}$ and $\vec{y}$.  If one assumes that the functional form of $A_{\lambda}(\cdot)$ and $B_{\lambda}(\cdot)$ changes in time (see also \cite{Hess} for the role of time in Bell's theorem) then the fixing of $P(\vec{a},\vec{b})=0$ can take place at times different than the measurement parameters selection and the sets in equations (\ref{e6}) and (\ref{e7}) will always be possible.


{\bf Received: Month 09, 2010}

\end{document}